\documentclass[12pt,preprint]{revtex4-1}

\usepackage{graphicx}
\usepackage{dcolumn}
\usepackage{natbib}

\usepackage{amsmath,amssymb}

\begin{document}

\title{Comment on "Velocity slip coefficients based on the hard-sphere 
Boltzmann equation" [Phys. Fluids 24, 022001 (2012)]}

\author{Silvia Lorenzani}
\affiliation{
Dipartimento di Matematica, Politecnico di Milano, 
Milano, Italy 20133}

%\begin{document}

%\maketitle

%\begin{abstract}

%\end{abstract}

\maketitle

%\clearpage
%\newpage

\section*{}

Recently, Gibelli published a paper \cite{1} about the derivation of the 
first- and
second-order velocity slip coefficients. 
In this paper, the author chooses
of comparing his results with those obtained by means of two variational
methods \cite{2}, \cite{3},  
underlying how his method predicts results
close enough to those reported in \cite{2}, while there is a 
"remarkable disagreement"
with those presented in \cite{3}, although the latter "compare favorably with 
the experimental data".
The aim of this comment is to try to trace back the origin of this discrepancy
considering that the two variational principles presented 
in \cite{2} and in \cite{3} are indeed deeply different.
The variational method used in \cite{2} applies to the integral
form of the linearized Boltzmann equation, which is available explicitly only
for simplified kinetic models.
In the more general case of the Boltzmann equation based on the true linearized
collision operator and general boundary conditions,
it is not possible
to obtain closed form expressions for all the different operators appearing
in the integral representation, which should consequently be approximated by
series expansions.
Therefore, in \cite{2}, the authors obtain approximate
solutions of an approximate integral equation where truncated series need to
be managed in order to obtain solutions of the Boltzmann equation in closed
form.
The last point (that is, the use of truncated series) is the common link with
the method of solution presented in \cite{1}, 
although within a completely different framework.
On the contrary, the variational technique presented  
in \cite{3} applies directly to the integrodifferential form of the
linearized Boltzmann equation and can be used for any linearized collision
term and extremely general boundary conditions.
This means that, in \cite{3}, the
right Boltzmann equation for hard-sphere molecules has been solved without
approximations in its form.
Only two eight-fold integrals have been evaluated numerically by using a
Monte Carlo integration, but since these two integrals are of the same simple
form of those analytically solved in \cite{4}, 
they could be computed with high accuracy.
All the rest has been carried out analytically.
The method of solution presented by
Gibelli requires the numerical evaluation of several eight-fold integrals but
in this case the statistical error in the Monte Carlo integration is not
negligible, as himself states (pag. $6$ in \cite{1}).
Concerning the role of the test function, there exist several basic theorems
which allow to perform a good choice.
In our context, then, this choice is even simpler since the functions which
need to be approximated by trial functions are solutions of physically 
realistic problems and expecially in their asymptotic forms can be immediately
obtained via the use of the Chapman-Enskog procedure, as done in \cite{2}, 
 or via the solution of the Boltzmann equation in integral
form based on a simplified kinetic model, as done in \cite{3}.
In \cite{1}, pag. $11$, the author writes:"... 
the current approach ... provides
the velocity slip coefficients with good accuracy in the entire range of the
accommodation coefficient".
It is difficult to imagine how he could deduce this good accuracy ...
The only explanation could be that he infers the accuracy of his asymptotic
near-continuum solution starting from the accuracy of his solution for the
Poiseuille mass flux valid in the whole range of the Knudsen numbers.
Unfortunately, this inference is completely wrong from the mathematical point
of view. 
This is proved by Gibelli himself.
Indeed, Fig. $1$ and Table II in \cite{1} show that the results concerning the 
complete
velocity and flux profiles obtained by Gibelli are in fair agreement with those
presented in \cite{5} for different Knudsen numbers and accommodations 
coefficients.
But then Fig. $3$  in  \cite{1}
shows clearly that the asymptotic solution for the Poiseuille
mass flux truncated at second order is completely wrong in the slip flow region
(where the second-order slip coefficient should be deduced).
A further example in this respect is offered by the BGK model.
The solution of the Poiseuille flow rate problem obtained by using the 
linearized Boltzmann equation based on the BGK model slightly deviates (only
a few percent) from that obtained by using the linearized Boltzmann equation
for hard-sphere molecules  \cite{6}, while the asymptotic 
second-order solutions 
differ considerably from each other, as pointed out in \cite{3}.
This is due to the fact that the BGK model is a 'first order' model.
Therefore, it can not be considered so accurate when higher order terms are
concerned, since a single relaxation time model can not give a good 
approximation for both first order and second order effects \cite{8}.
The method of solution presented in \cite{1} suffers from the same problem,
although within a completely different framework.
In fact, even if Gibelli starts from the exact hard-sphere Boltzmann equation,
the method of solution that he proposes consists in expanding the distribution
function in terms of half-range Hermite polynomials and then a system of moment
equations in the expansion coefficients is solved so as to "satisfy 
approximately the hard-sphere Boltzmann equation", as himself states.
Therefore, in order to find closed form solutions, he has to truncate the
polynomial expansion of the distribution function at some finite order. 
The final result of the procedure presented by Gibelli in \cite{1} 
is summarized in
Fig. 3 of his paper, where it is clearly shown that the
second-order description of the flow rate $Q$ is completely incorrect in the
slip region.
To overcome this difficulty Gibelli suggests to introduce
a third slip
coefficient, $A_3$, and presents the following expansion for 
the volume flow rate:

$$
\frac{Q}{K} \simeq \frac{1}{12 \epsilon \sqrt{\pi}} \bigg(\frac{A_o}{K_n}+
6 \overline{A}_1+12 K_n \overline{A}_2-12 {K_n}^2 \overline{A}_3 \bigg) \; \;
\; \; \;  \; \; \;  \; \; \; \; \; \;                (27)
$$
(where the bar is used to distinguish the velocity slip coefficients which
account for the structure of the Knudsen layer). Then he writes: "the
volume flow rate given by Eq. ($27$) with $A_o=1$, in fact, can also be obtained
by integrating the velocity profile satisfying the linearized Navier-Stokes
equation, subject to the boundary conditions" at the walls:

$$
\xi_x=\mp \overline{A}_1 K_n \frac{d \xi_x}{dy}-{K_n}^2 (\overline{A}_2-
\overline{A}_3 K_n) \frac{d^2 \xi_x}{dy^2}  \;  \; \; \; \;
\; \; \;  \; \; \;  \; \; \; \; \; \; \; \; \; \;  \; \; \; \;       (28)
$$
where $y$ is the coordinate normal to the walls, $\xi_x$ is the tangential
velocity component and $K_n$ is the Knudsen number.
Unfortunately, this inference is incorrect in several respects.
\begin{itemize}
\item{} From the mathematical point of view, since the slip coefficients
which multiply each normal derivative of the velocity profile in the
boundary condition ($28$) should have an order of accuracy in the Knudsen
number as the order of the derivative itself, that is $A_2$ is a second-order
slip coefficient and should multiply the second derivative of $\xi_x$, while
$A_3$ is a third order slip coefficient and should multiply the third
derivative of $\xi_x$.
This is easily deduced by dimensional reasons and is clearly proved in 
\cite{8}, \cite{10}. 
But in Eq. ($28$) in front of the second derivative of $\xi_x$
appears a strange mix of
coefficients of different order in the Knudsen number. 
\item {} From a physical point of view, since the third derivative of the
Poiseuille velocity profile is zero, therefore third order effects in the flow
rate $Q$ (included in $A_3$) are not expected to arise from the third order
slip coefficient (which is the coefficient in front of 
$\frac{d}{dy}(\frac{d^2 \xi_x}{dy^2})$) but from the contribution of the
Knudsen layer \cite{8}.
\item {} From a conceptual point of view, since Gibelli writes the 
boundary conditions to be associated to the Navier-Stokes equation (Eq. ($28$))
in terms of the 'effective' slip coefficients (that is, slip coefficients
which account for the structure of the Knudsen layer).
Instead, a successful higher-order slip model is one that does not agree with
the Boltzmann equation solutions in the Knudsen layers, in order to be able
to predict within the same framework all the physical quantities of interest:
not only the flow rate but also, for example, the stress field which is not
subject to a Knudsen layer correction.
\end{itemize}
Beyond these flaws, the major shortcoming of the work presented by Gibelli is
that this author has avoided to compare his results with the
outcomes obtained in some recent experimental studies. 
Indeed, the values of the slip coefficients determined in two independent
series of experiments carried out for different gases  
\cite{12}, \cite{14}
show a very good agreement with those obtained in
\cite{3}, while there is a remarkable disagreement with the results 
presented  in \cite{1} concerning the second-order slip coefficient.

\end{document}